\documentclass{article}
\begin{document}

\title{Dark matter and dark energy proposals:\\
maintaining cosmology as a true science? \footnote{Paper for
CRAL-IPNL conference "Dark Energy and Dark Matter", Lyon 2008.}}

\author{George F R Ellis \footnote{Mathematics Department, University of Cape
Town.}}
\maketitle

\begin{abstract} I consider the relation of explanations
for the observed data to testability in the following contexts:
observational and experimental detection of dark matter;
observational and experimental detection of dark energy or a
cosmological constant $\Lambda$; observational or experimental
testing of the multiverse proposal to explain a small non-zero
value of $\Lambda$; and observational testing of the possibility
of large scale spatial inhomogeneity with zero $\Lambda$.
\end{abstract}
%


\section{Dark matter and testability}
As discussed at this meeting, there is a great deal of
astronomical evidence for dark matter: galaxy rotation curves and
dynamical studies; the Baryon Acoustic Oscillations (BAO) and
peaks in the Cosmic Blackbody Radiation (CBR) power spectrum; and
other studies of Large Scale Structure (LSS). An important feature
of this story is that dark matter was (somewhat reluctantly)
observationally discovered, it was not predicted! However its
nature is unknown, except that it is not baryonic, so there is
much theoretical speculation about what it is.

As to the astronomical evidence, there is more coming, with many
studies under way and detectors planned. Laboratory and accelerator
tests are also planned, attempting to link astrophysics to detected
particle properties. This will be a major coup if successful - it
will identify dark matter physically. Accompanying this experimental
work is a continuing development of different theories about the
nature of dark matter, as evidenced at this meeting.

All this work is very much in the proper scientific spirit: make
theories and test observationally and experimentally. Dark matter
searches and tests are thriving, and I will not comment more on them
here.

\section{The Acceleration of the universe}
The case of dark energy is quite different. The explanation of dark
energy is a central preoccupation of present day cosmology. Its
presence is indicated by the recent speeding up of the expansion of
the universe as shown by the supernova observations discussed at
this meeting, and is confirmed by other observations such as those
of the cosmic background radiation anisotropies and LSS/BAO studies.
Like dark matter, its existence was discovered, not predicted.

As discussed at this meeting, astronomical observations are being
refined in many sophisticated ways and used to confirm the
acceleration and test equations of state of hypothetical dark
energy. However the interpretation of these observations is
ambiguous, as discussed in section 4. It is therefore crucial to
pursue the possibility of any other tests on the one hand, and
theoretical explanations on the other. So how can we confirm its
existence and nature?

\subsection{Lab tests of Dark energy? }
It is striking that at this meeting there are many proposals for
laboratory or accelerator detection of dark matter, but none for
dark energy. Indeed such tests in a lab or even the solar system are
not feasible, in the case of the usual conception of DE as
cosmological constant or quintessence; it simply has no significant
effect at the relevant scales. The exception would be in the case of
unified approaches to Dark Energy and Dark Matter, as sometimes
discussed.

Such approaches need to be explored: they may be facets of the same
problem, and then evidence for dark matter is also evidence for dark
energy. But then we would have a force that would change (with
scale) from attraction to repulsion: we would have to explain why
and how, and propose how to test that change. The lab tests for dark
matter would not explore the scales where it would become effective
dark energy. It seems unlikely one would attain the required
evidence for the dark energy effect.

\subsection{Theoretical explanations?}
Without lab tests, we have to rely on theoretical explanations for
its nature. However that nature (whether constant, or varying) is a
major problem for theoretical physics. It is not uniquely related to
any known field or particles.

If the dark energy is in fact constant, the attempt to explain it
from fundamental physics is a disaster --- theoretical proposals for
a cosmological constant from quantum field theory give an answer
$10^{120}$ factors too large! The only way out seems to be the
multiverse proposal, which is gaining ground, but is rather
problematic as a scientific explanation, as discussed in section 3.

If it is varying, a quintessence  field, we need to know its nature;
but no compelling identification has been made. Perhaps it is due to
modified gravitational theories: higher curvature terms or effects
of higher dimensions. Perhaps it is due to some physical effect such
as Bose-Einstein condensation; such options need to be explored. We
can just deal with it at a phenomenological level in terms of an
arbitrary equation of state; but not all such equations of state are
physically acceptable, as discussed in Section 5.4, and in any case
that approach is not in the end satisfactory: ultimately it needs an
underlying physical basis.

In all cases, the issue is how do we test these theoretical
proposals? Many seem very arbitrary. Just writing down a Lagrangian
does not prove such matter exists! If the explanation only explains
one thing (the observed acceleration) and has no other testable
outcome, it is an ad hoc explanation for that one thing rather than
a unifying scientific proposal. It needs some other independent
experimental or observational test – but we don't have another
viable context for applying such tests.

So how do we justify our proposed  theoretical explanations? Why
this form of quintessence? Why a cosmological constant with the
observed value? We need to see if there are any alternatives; and
there are, as discussed below.

\section{Explaining the cosmological constant via a multiverse}
The idea of a multiverse -- an ensemble of universes or of universe
domains – has received increasing attention in cosmology (see the
articles in Carr \cite{car07} for an up to date survey), with
suggestions including that it can occur

- in separate places, as particularly justified by chaotic inflation
(Linde \cite{linetal94,lin03}, Guth \cite{gut01,gut07}, Vilenkin
\cite{vil06})

- through the Everett quantum theory interpretation: other
branches of the wavefunction of the universe (Deutsch
\cite{deu97})

- because of the landscape of string theory, imbedded in a chaotic
cosmology (Susskind \cite{sus05}).\\

\noindent A particular theoretical driver of these proposals is
the "anthropic" issue : the realization that the universe is
fine-tuned for life as regards both the laws of physics and as
regards the boundary conditions of the universe (Barrow and Tipler
\cite{bartip86},  Rees \cite{ree99,ree01}). A multiverse with
varied local physical properties is  one possible scientific
explanation: an infinite set of universe domains allows all
possibilities to occur, so somewhere things work out OK:
conditions for life will be fulfilled somewhere in the multiverse
(NB: it must be an actually existing multiverse rather than a
hypothetical one - this is essential for any such anthropic
argument).

The application of this proposal is to explaining fundamental
constants, and particularly explaining the small value of the
cosmological constant (Weinberg \cite{wei00,wei07}, Susskind
\cite{sus05}). Too large a value for $\Lambda$ results in no
structure and hence no life; so anthropic considerations in a
multiverse mean that the value of $\Lambda$ observed by any
intelligent being will be small (in fundamental units), thus
justifying an actual value extremely different from the `natural'
one predicted by physics: a difference of $120$ orders of magnitude.
This makes clear the true multiverse project: making the extremely
improbable appear probable.

However the very nature of the scientific enterprise is at stake in
the multiverse debate: the multiverse proponents are proposing
weakening the nature of scientific proof in order to claim that
multiverses provide a scientific explanation. This is a dangerous
tactic (note that we are concerned with really existing multiverses,
not potential or hypothetical). Two central scientific virtues are
testability and explanatory power. In the cosmological context,
these are often in conflict with each other (Ellis \cite{ell06}).
The extreme case is multiverse proposals, where no direct
observational tests of the hypothesis are possible, as the supposed
other universes cannot be seen by any observations whatever, and the
assumed underlying physics is also untested and indeed probably
untestable.

In this context one must re-evaluate what the core of science is:
can one maintain one has a genuine scientific theory when direct and
indeed indirect tests of the theory are impossible? If one claims
this, one is altering what one means by science. One should be very
careful before so doing. The key observational point is that the
domains considered are beyond the particle horizon and are therefore
unobservable. The assumption is we that can extrapolate to 100
Hubble radii, $10^{1000}$ Hubble radii, or much much more (infinity
is often mentioned); but we have no data whatever about these
domains. Given this extremely poor observational context, are there
other
reasons to believe the multiverse proposal?\\

\emph{Is it implied by known physics, that leads to chaotic
inflation?}

\noindent The key physics (Coleman-de Luccia tunneling, the string
theory landscape) is extrapolated from known and tested physics to
new contexts; the extrapolation is unverified and indeed is
unverifiable; it may or may not be true. The physics is hypothetical
rather than tested. Is the situation:\\

            Known Physics   $\Rightarrow$    Multiverse ??\\

\noindent NO! The real situation is\\

Known Physics  $\Rightarrow$ ? Hypothetical  Physics $\Rightarrow$
Multiverse\\

\noindent It is a great extrapolation from known physics. This
extrapolation is untested, and may be untestable:  it may or may not be correct.\\

\emph{Is it Implied by inflation, which is justified by CBR
anisotropy observations?}

\noindent It is implied by some forms of inflation but not others;
inflation is not yet a well defined theory, it is a family of
theories.
Not all forms of inflation lead to chaotic inflation, for example
inflation can occur in small closed universes.\\

\emph{Is it implied by probability arguments?}

\noindent It is claimed it is implied by a probability argument: the
universe is no more special than need be to create life. Hence the
observed value of the Cosmological constant is confirmation
(Weinberg \cite{wei00,wei07}; Rees \cite{ree99,ree01}). But the
statistical argument only applies if a multiverse exists; it is
simply inapplicable if there is no multiverse. In that case we only
have one object we can observe; we can do many observations of that
one object, but it is still only one object (one universe), and you
can't do statistical tests if there is only one existent entity
(Ellis \cite{ell06}). Furthermore, we don't know the measure to use;
but the result depends critically on this choice. Overall, this is a
weak consistency test on multiverses, that is indicative but not
conclusive, firstly because a probability argument cannot in fact be
falsified, all it can do is confirm that some result is improbable;
and secondly, because while consistency tests must be
satisfied, they are not confirmation unless no other explanation is possible.\\

\emph{Is it testable through predicting closed spatial sections?}

\noindent The claim is made (Susskind \cite{sus05}) that only
negatively curved RW models can emerge in a chaotic inflation
multiverse, because Coleman-de Luccia tunneling only gives such
models; so one can disprove chaotic inflation if one
observationally determines that $k=+1$. But that claim about
inflation is already disputed, as there are papers suggesting
$k=+1$ tunneling is possible. In any case this model it depends on
a very specific speculative mechanism, which has not been verified
to actually work, and indeed such verification is probably
impossible. Alternatively one can claim the idea is disproved if
we determine the spatial sections are positively curved in the
observed region, for then if they extend unchanged far enough this
implies a closed universe and hence no chaotic inflation. But we
could live in high density lump imbedded in a low density
universe: the extrapolation of $k=+1$ geometry beyond the visual
horizon may not be valid. Neither argument is conclusive!

However, chaotic inflation \emph{can} be disproved if we
observationally prove we live in a \emph{small universe}: that is,
we have already seen round the universe because it has small closed
spatial sections. To test for this possibility we can search for
identical circles in the CBR sky, plus a low CBR anisotropy power at
large angular scales (which is what is observed) -- see Frank Steiner's
contribution. This is an
important test as it would indeed disprove the chaotic inflation
variety of multiverse; but not seeing them would not prove a
multiverse exists. Their non-existence is a necessary
but not sufficient condition for a multiverse.\\

\emph{Is it theoretically preferable to other explanations of the
way things are?}

\noindent This is what many are claiming. Indeed the real argument
for the proposal is that it is the only purely physical explanation
for fine tuning of parameters that lead to our existence, in
particular the value of the cosmological constant; but this is
theoretical explanation, not supported by astronomical observation.
So which is more important in cosmology: theory (explanation) or
observations (tests against reality)? That is the core issue to be faced.\\

\emph{Is it an infinity of entities necessarily implied?}

Often it is claimed there are physically existing infinities both of
universes in a multiverse, and of spatial sections in each of the
universes in the multiverse context, see e.g. Vilenkin \cite{vil06}.
But infinity is an unattainable state rather than a number, and is
plausibly never attained in physical reality. Indeed David Hilbert
 states \emph{"the infinite is nowhere to be found in
reality, no matter what experiences, observations, and knowledge are
appealed to"} (Hilbert \cite{hil64}). Furthermore, this claim is
completely untestable: if we could see them, which we can't, we
could not count them in a finite time! The claimed existence of
physically existing infinities is highly dubious, and is not a
scientific statement, if science involves testability by either
observation or experiment. This claim in the multiverse context
emphasizes how tenuously scientific that
idea is. It is not remotely testable.\\

\emph{Implication of all the above: }

\noindent The multiverse idea is not provable either by observation,
or as an implication of well established physics (cf. Gardner
\cite{gar03}). It may be true, but cannot be shown to be true by
observation or experiment. However it does have great explanatory
power: it does provide an empirically based rationalization for fine
tuning, developing from known physical principles. Here one must
distinguish between explanation and prediction. Successful
scientific theories make predictions, which can then be tested. The
multiverse theory can't make any predictions because it can explain
anything at all. Any theory that is so flexible is not testable
because almost any observation can be accommodated. I conclude that
multiverse proposals are good empirically-based philosophical
proposals for the nature of what exists, but are not strictly within
the domain of science because they are not testable. I emphasize
that there is nothing wrong with empirically-based philosophical
explanation, indeed it is of great value, provided it is labeled for
what it is. I suggest that cosmologists should be very careful not
to make methodological proposals that erode the essential nature of
science in their enthusiasm to support such theories as being
scientific (cf. Tegmark \cite{teg03,teg04}), for if they do so,
there will very likely be unintended consequences in other areas
where the boundaries of science are in dispute. It is dangerous to
weaken the grounds of scientific proof in order to include
multiverses under the mantle of `tested science' for there are many
other theories standing in the wings that would also like to claim
that mantle.

\section{Inhomogeneity and the Acceleration of the universe}

The deduction of the existence of dark energy is based on the
assumption that the universe has a Robertson-Walker (RW) geometry
- spatially homogeneous and isotropic on a large scale. The
observations can at least in principle be accounted for without
the presence of any dark energy, if we consider the possibility of
inhomogeneity. This can happen in two ways: locally via
backreaction and observational effects, and via large scale
inhomogeneity.

\subsection{Small scale inhomogeneity: backreaction and observational effects}
Acceleration due to back reaction from "small scale" inhomogeneities
is discussed by Wiltshire at this meeting (see also Wiltshire
\cite{wil08}, Buchert \cite{buc08}). Wiltshire proposes that
gravitational energy can provide a source of effective dark energy,
leading to the possibility in principle of concordance cosmology
without $\Lambda$. It is important to notice there are two effects
here: firstly the backreaction from small scale inhomogeneity to the
large scale geometry can generate a dynamic effect in the effective
Friedmann equation for the cosmology, and secondly small scale
inhomogeneity has significant effects on the propagation of photons
in a lumpy universe, with potentially important effects on
observations.

Whether these effects are sufficient to account for the apparent
supernova observations is an important ongoing debate involving
interesting modeling and general relativity issues, and particularly
how one models a universe with large scale voids and the nature of
the Newtonian limit in cosmology. In my view the jury is still out
on this one, with many skeptical there is any significant effect and
others suggesting it may be at least large enough to affect the
cosmic relation between energy densities and expansion that leads us
to deduce the spatial curvature is almost flat. Conceptual clarity
on the modeling issues involved is required.

\subsection{Large scale inhomogeneity: inhomogeneous geometry}

Perhaps there is a large scale inhomogeneity of the observable
universe such as that described by the Lema\^{\i}tre-Tolman-Bondi
(LTB) pressure-free spherically symmetric models, and we are near
the centre of a void. The idea that such models can explain the
supernova observations without any dark energy is discussed by
Cel\'{e}ri\'{e}r at this meeting (and see also C\'{e}l\'{e}rier
\cite{cel07}).

\subsection{Can we fit the observations?}
The LTB models have comoving coordinates

$$ds^2 = - dt^2 + B^2(r,t) + A^2(r,t)(d\theta^2+\sin^2 \theta
d\phi^2)$$ where

$$B^2(r,t) = A'(r,t)^2 (1-k(r))^{-1}$$
and the evolution equation is

$$(\dot{A}/A)^2 = F(r)/A^3 + 8pG\rho_\Lambda/3 - k(r)/A^2$$
with the energy density given by $F' (A'A^2)^{-1} = 8pG\rho_M$.

There are two arbitrary functions of the spatial coordinate $r$:
namely $k(r)$ (curvature) and $F(r)$ (matter). This freedom enables
us to fit the supernova observations with no dark energy or other
exotic physics (this is a theorem, see Mustapha \emph{et al}
\cite{mustetal99}). One can also fit the CBR observations because
they refer to much larger values of $r$ (see e.g. Alexander \emph{et
al} \cite{aletal07}). One should note here that at least some of the
observed CBR dipole can then arise because we are a bit off-centre
in the void, so one can re-evaluate the great attractor analysis in
this context and the alignment of the dipole and quadrupole.
Nucleosynthesis data can also be fitted; what is a bit more
problematic is the BAO. The key comment to make is that different
scales are probed by different observations and can in principle all
be fitted by adjusting the free spatial functions at different
distances.

A typical observationally viable model is one in which we live
roughly centrally (within 10\% of the central position) in a large
void: a compensated underdense region stretching to $z \simeq 0.08$
with $\delta \rho/\rho \simeq -0.4$ and size $160/h$ Mpc to $250/h$
Mpc, a jump in the Hubble constant of about $1.20$ at that distance,
and no dark energy or quintessence field (Biswas \emph{et al}
\cite{bisetal06}, Ishak \emph{et al} \cite{ishetal07}, Yoo \emph{et
al} \cite{yooetal08}).

\subsection{Large scale inhomogeneity: dynamic evolution?}

Given we can fit the observations by such a model, can we find
dynamics (inflation followed by a HBB era) that can lead to such a
model? It has the same basic dynamics as the standard model
(evolution along individual world lines governed by the Friedmann
equation) but with distant dependent parameters. Will inflation
prevent it? This depends on the initial data, the amount of
inflation, and the details of the unknown inflaton. If we are
allowed the usual tricks of fiddling the inflationary potential and
initial data, and adding in multiple fields as desired, then there
is sufficient flexibility that it should certainly be possible.

\subsection{Improbability}
Many dismiss these models on probability grounds: "It is
improbable we are near the centre of such a model." But there is
always improbability in cosmology. We can can shift it around, but
it is always there. It might be in the nature of a
Robertson-Walker geometry (the old view), in the inflationary
potential  and initial conditions (the current mainstream
position), which specific universe domain we are in within a
multiverse, or the spatial position in an inhomogeneous universe
(the present proposal). Note that we are competing with a
probability of $10^{-120}$ for $\Lambda$ in a RW universe; we do
not have to get very high probabilities to outdo that
improbability, which is what the multiverse proposal aims to
handle.

Three comments are in order. First, a key feature of cosmology is
that there is only one universe; and the very concept of probability
does not apply to a single object, even though we can make many
measurements of that single object to determine its detailed nature.
Probability applies to the multiple measures we can make of the
single universe, but not to issues doing with the existence of the
universe itself (Ellis \cite{ell06}). There is no physically
realised ensemble to apply that probability to, unless a multiverse
exists in physical reality – which is not proven, as discussed
above: it's a philosophical assumption. In essence, there simply is
no proof the universe is probable; that is a philosophical
assumption, which may not be true. The universe may be improbable!!
Secondly, there is no well-justified measure for any such
probability proposal even if we ignore the first problem. This is
still an issue of debate.

And thirdly, a study by Linde \emph{et al} \cite{linetal95} shows
that (given a particular choice of measure) this kind of
inhomogeneity actually is a probable outcome of inflationary theory,
with ourselves being located near the centre!  One cannot dismiss
such models out of hand for probability reasons.

\section{Observational tests of spatial homogeneity}
Given the above context, direct observational tests of the
Copernican(spatial homogeneity) assumption are of considerable
importance. Given that we can both find inhomogeneous models to
reproduce the observations without any exotic energy, as well as
homogeneous models with some form of dark energy that explain the
same observations, can we distinguish between the two? Ideally we
need a model-independent test: is a RW geometry the correct metric
for the observed universe region? Four kinds of tests are
possible, as discussed below. Whatever position we may have on the
issue of probability, in the end our philosophy on this question
will have to give way to any such possible observational tests.

\subsection{CBR based tests}
Some tests use scattered CBR photons to check spatial homogeneity
(Goodman \cite{goo95}; Caldwell and Stebbins \cite{calste07}). If
the CBR radiation is anisotropic around distant observers (as will
be true in inhomogeneous models), Sunyaev-Zeldovich scattered
photons have a distorted spectrum that reflects the spatial
inhomogeneity. However this test is somewhat model dependent - it
is good for void models but misses, e.g., conformally stationary
spacetimes. It also has to take into account other possible causes
of spectral distortion.

\subsection{Direct observational tests: behaviour near origin}
The universe must not have a geometric cusp at the origin, as this
implies a singularity there. Thus there are centrality conditions
that must be fulfilled in the inhomogeneous models (Vanderveld et al
\cite{vanetal06}). The distance modulus behaves as $\Delta dm(z) = -
(5/2)q_0z$ in standard $\Lambda$CDM models, but if this were true in
a LTB void model without $\Lambda$ this implies a singularity
(Clifton \emph{et al} \cite{clietal08}). Observational tests of this
requirement will be available from intermediate redshift supernovae
in the future.

\subsection{Direct observational tests: constancy of curvature}
There are two geometric effects on distance measurements:
curvature $\Omega_k$ bends null geodesics, expansion $H(z)$
changes radial distances. These are coupled in RW models, as
expressed in the relation
$$ d_L(z)= \frac{(1+z)}{H_0\sqrt{-\Omega_k}}
\sin{\left(\sqrt{-\Omega_k}\int_0^z dz'\frac{H_0}{H(z')}\right)}.$$
While these effects are strictly coupled in RW geometries, they are
decoupled in LTB geometries.

In RW geometries, we can combine the Hubble rate and distance data
to find the curvature today:
$$\Omega_k = \frac{[H(z)D'(z)]^2-1}{H_0D(z)]^2} $$
This relation is independent of all other cosmological parameters,
including dark energy model and theory of gravity. It can be used at
single redshift to determine $\Omega_k$. The exciting result of
Clarkson \emph{et al} (\cite{claetal08}) is that since $\Omega_k$ is
independent of $z$, we can differentiate to get the consistency
relation
$$ C(z) := 1 + H^2(DD''-D'^2) + HH'DD' =0, $$
which depends only on a RW geometry: it is independent of
curvature, dark energy, nature of matter, and theory of gravity.
Thus it gives the desired consistency test for spatial
homogeneity. In realistic models we should expect $C(z) \simeq
10^{-5}$, reflecting perturbations about the RW model related to
structure formation. Errors may be estimated from a series
expansion

$$C(z) = \left[q_0^{(D)}-q_0^{(H)}\right]z + O(z^2)$$
where $q_0^{(D)}$ is measured from distance data and $q_0^{(H)}$
from the Hubble parameter. It is simplest to measure $H(z)$ from
BAO data. It is only as difficult carrying out this test as
carrying out dark energy measurements of $w(z)$ from Hubble data,
which requires $H'(z)$ from distance measurements or the second
derivative $D''(z)$.

This is the simplest direct test of spatial homogeneity, and its
implementation should be regarded as a high priority: for if it
confirms spatial homogeneity, that reinforces the evidence for the
standard view in a satisfying way; but if it does not, it has the
possibility of undermining the entire project of searching for a
physical form of dark energy.

\subsection{Indirect Observational tests}
If the standard inverse analysis of the supernova data to
determine the required equation of state, as discussed at this
meeting, shows there is any redshift range where $w := p/\rho <
-1$, this may well be a strong indication that one of these
geometric explanations is preferable to the Copernican
(Robertson-Walker) assumption, for otherwise the matter model
indicated by these observations is non-physical (it has a negative
kinetic energy).

There is already data suggesting this may be be the case, see e.g.
Lima \emph{et al} \cite{limetal08}. There are some attempts to
generate matter models that will give this kind of behaviour without
negative kinetic energies, but they are very speculative physically,
supposing multiple unknown forms of matter or energy with
arbitrarily proposed interactions between them. It all seems rather
reminiscent of the Ptolemaic epicycles for describing the solar
system. The physically most conservative approach is to assume no
unusual dark energy or exotic interacting fields, but rather that an
inhomogeneous geometry might be responsible for the observed
apparent acceleration; this should be seriously considered as an
alternative.

\section{Conclusion}
The issue of what is testable and what is not testable in
cosmology is a key issue. Some dark energy proposals, specifically
from multiverse advocates, propose weakening the link to
observational tests, because they believe we have such a good
theory that it must be right. But if a proposal is not testable,
we certainly need to consider observationally testable
alternatives.

The acceleration indicated by supernova data could possibly be due
to small scale inhomogeneity that definitely exists, but may not be
sufficiently significant to do the job. It could be due to large
scale inhomogeneity that can probably do the job, but may not exist.
Observational tests of the latter possibility are as important as
pursuing the dark energy (exotic physics) option in a homogeneous
universe. Theoretical prejudices as to the universe's geometry, and
our place in it, must bow to such observational tests.

We should stand firm and insist that genuine science is based on
observational testing of plausible hypotheses. There is nothing
wrong with physically motivated philosophical explanation: but it
must be labeled for what it is. Overall: theory must be subject to
experimental and/or observational test; this is the central feature
of science. There is good progress in this respect as regards both
dark matter and dark energy


\end{document}